\documentclass[conference]{IEEEtran}
\IEEEoverridecommandlockouts
\pdfoutput=1
\usepackage[noadjust]{cite}
\usepackage{amsmath}
\usepackage{breqn}
\usepackage{array}
\usepackage{mdwmath}
\usepackage{amssymb}
\usepackage{algorithm}
\usepackage[noend]{algpseudocode}
\usepackage{eqparbox}
\usepackage{stfloats}
\usepackage{url}
\usepackage{tikz}
\usepackage{pgfplots}
\graphicspath{{../pdf/}{../jpeg/}}
\DeclareGraphicsExtensions{.pdf,.jpeg,.png}
\usepackage{epstopdf}
\usepackage{amsfonts}
\usepackage{tikz}
\usepackage{ctable}

\newtheorem{theorem}{Theorem}

\usepackage{stackengine}
\def\delequal{\mathrel{\ensurestackMath{\stackon[1pt]{=}{\scriptstyle\Delta}}}}

\def\BibTeX{{\rm B\kern-.05em{\sc i\kern-.025em b}\kern-.08em T\kern-.1667em\lower.7ex\hbox{E}\kern-.125emX}}

\begin{document}

\title{\huge RIS-Aided Cell-Free Massive MIMO: Performance Analysis and Competitiveness}   

\author{\IEEEauthorblockN{Bayan Al-Nahhas, Mohanad Obeed, Anas  Chaaban, and Md. Jahangir Hossain}
\IEEEauthorblockA{School of Engineering, University of British Columbia, Kelowna, Canada. \\ 
Email:\{bayan.alnahhas, mohanad.obeed, anas.chaaban, jahangir.hossain\}@ubc.ca}}

\maketitle

\begin{abstract}

In this paper, we consider and study a cell-free massive MIMO (CF-mMIMO) system aided with reconfigurable intelligent surfaces (RISs), where a large number of access points (APs) cooperate to serve a smaller number of users with the help of RIS technology. We consider imperfect channel state information (CSI), where each AP uses the local channel estimates obtained from the uplink pilots and applies conjugate beamforming for downlink data transmission. Additionally, we consider random beamforming at the RIS during both training and data transmission phases. This allows us to eliminate the need of estimating each RIS assisted link, which has been proven to be a challenging task in literature. We then derive a closed-form expression for the achievable rate and use it to evaluate the system's performance supported with numerical results. We show that the RIS provided array gain improves the system's coverage, and provides nearly a 2-fold increase in the minimum rate and a 1.5-fold increase in the per-user throughput. We also use the results to provide preliminary insights on the number of RISs that need to be used to replace an AP, while achieving similar performance as a typical CF-mMIMO system with dense AP deployment. 

\end{abstract}

\vspace{-.06in}
\section{Introduction}
\vspace{-.02in}

\label{Sec:Intro}

Massive connectivity and demanding coverage requirements in the emerging sixth generation (6G) wireless networks brought in the need of new and revolutionary technologies for wireless communication systems. Two technological advances have recently caught significant attention in literature and are expected to take an important part in the future 6G wireless networks, namely the cell-free massive multiple input multiple output (CF-mMIMO) systems and reconfigurable intelligent surfaces (RISs). CF-mMIMO systems are considered as a promising technology due to their architecture advantage, which is seen to reap the benefits of massive MIMO systems and distributed systems together. CF-mMIMO systems can be viewed as a particular deployment of mMIMO systems, where the network consists of a large number of APs spread out over a certain georgraphical area. In CF-mMIMO, the boundaries among cells are removed and all the APs cooperate to serve all the users, thus the name. The distributed architecture allows the CF-mMIMO systems to fully exploit macrodiversity and potentially offers high probability of coverage. The reason is that the users in CF-mMIMO are close to APs, which improves the systems fairness by enhancing the service to edge-user \cite{Ngo,Zhang}.\\
Many recent results on CF-networks show that CF-networks outperfoms the conventional cellular and small cell networks under various practical settings \cite{Nayebi,Ngo, QNgo}. Recent works also studied CF-mMIMO asymptotic performance under channel hardening and favorable propagation assumptions \cite{Chen, Chen2}. Nevertheless, like many technologies, CF-mMIMO also suffers from drawbacks including backhauling traffic and infrastracture cost \cite{AAIB}.\\
Recently, the concept of deploying RISs in existing communication systems has emerged as a cost effective solution. The RIS is envisioned as a planar array of passive reflecting elements that can independently induce phase-shifts onto the incident electromagnetic waves for performance enhancement \cite{haung,Marco}. Several papers have demonstrated the effectiveness of RISs in envisioning a smart and interconnected enviroment for future wireless generations \cite{Gong,pan}. RISs have also been studied in existing wireless technologies, bringing about names such as RIS-aided massive MIMO systems, RIS-aided NOMA systems, RIS-aided security system and RIS-aided UAV communication \cite{bayan, nadeem, secure, noma, UAV}. All these studies provide insightful analysis on the improved performance while envisioning lower cost and higher efficiency than existing systems. Recently, RISs have also been introduced to CF networks and studied under various setups showing potential effectiveness \cite{2020decentralized, 2020capacity, 2020energy}.

Motivated by the above, this paper considers an RIS-aided CF-mMIMO (RIS-CF-mMIMO) system and studies the resulting performance enhancement. Specifically, we study a multi-user downlink scenario in a CF-mMIMO system, wherein transmission is aided by a number of RISs uniformly distributed in a geographical area. Unlike previous works on RIS-CF networks, we assume an imperfect channel state information (CSI) at the AP and derive a closed-form acheivable rate expression while considering conjugate beamforming at the APs. We show that in a Rayleigh fading environment and using the direct etimation (DE) protocol proposed in \cite{bayan}, RISs provide an array gain dependent on the number of elements, but their phases do not play a significant role in improving signal-to-interference-plus-noise ratio (SINR) at the user end. Since a well known advantage of CF-mMIMO systems is providing competitive coverage performance when compared to small cells, we study the additional effect of integrating RISs configured with random phase-shifts in further enhancing coverage. Similar to previous works, we adopt the theoretical minimum rate and average per user throughput, which accounts for pilot overhead, as our perfomance criteria. Specifically, we consider a $5\%-20\%$ outage probability among all users for system analysis \cite{Nayebi}. We extend the analysis to study the acheivable sum rate of a CF-mMIMO system with a number of APs and RISs, and compare it with a typical CF-mMIMO system with a dense AP deployment. Numerical results show promising preliminary insights on the potential of deploying RIS in the current CF-mMIMO topology.

The rest of the paper is organized as follows: Sec. II describes the system model. Sec III explains the channel estimation. Sec. IV investigates the conjugate beamforming and provides the acheivable rate expression. Sec. V provides the numerical results and Sec. VI concludes the paper.  
\begin{figure}[!t]
\centering
\includegraphics[width=.5\textwidth, height=.3\textwidth]{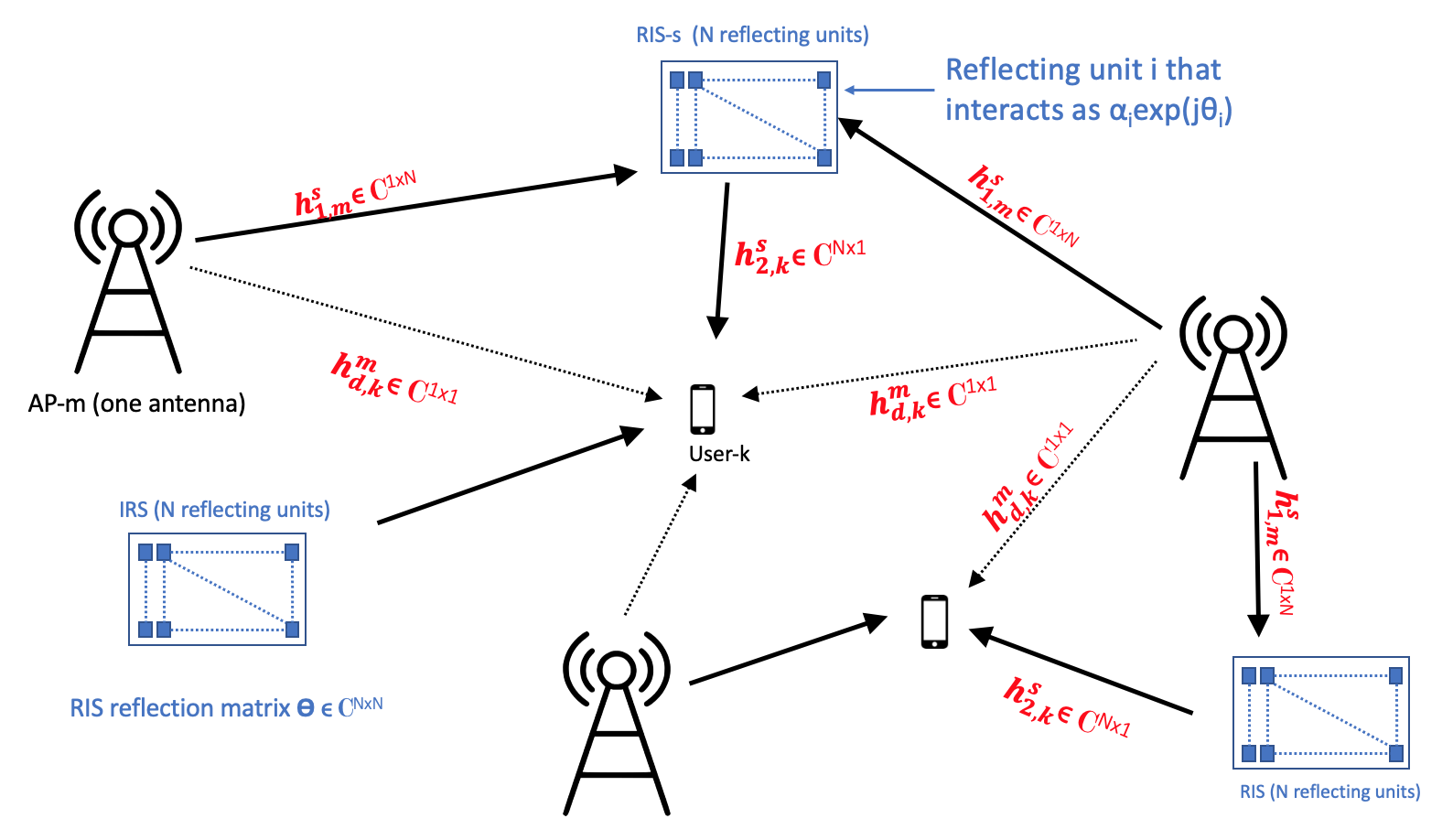}
\caption{Sketch of an RIS-CF-mMIMO system.}
\label{LIS_model}
\end{figure}

\section{System Model}

As shown in Fig. \ref{LIS_model}, we consider a system consisting of $M$ single-antenna APs communicating with $K$ single-antenna users. The system is assisted by $S$ RISs, each composed of $N$ passive reflecting elements installed in the line-of-sight (LoS) of the APs. In this topology, we assume that the APs, users, and RISs are uniformly distributed in $D \times D$ $\text{km}^{2}$ geographical area, where $S,M \gg K$. Furthemore, we consider a time division duplex (TDD) protocol under which the APs utilize the uplink training phase to obtain the local channel estimates and the downlink transmission phase to apply conjugate beamforming and transmit data. Our objective is to study the downlink acheivable rate at the user end for the proposed IRS-CF-mMIMO system and derive its closed-form expression for performance analysis. In the following, we will define channels depicted in Fig. \ref{LIS_model} and express the signal model of our proposed system.

The LoS channel between the $m$'th AP and $s$'th RIS is denoted as $\mathbf{h}_{1,m}^{s}=\sqrt{\beta_{1,m}^{s}}[{h}_{1,m,1}^{s}, \dots , {h}_{1,m,N}^{s}] \in \mathbb{C}^{1\times N}$, where $\beta_{1,m}^{s}$ is the channel attenuation factor and ${h}_{1,m,n}^{s}$ is the small-scale fading coefficient between the $m$'th AP and the $n$'th reflecting element in the $s$'th RIS. Also, $\mathbf{h}_{2,k}^{s}\sim \mathcal{CN}(\mathbf{0},\beta_{2,k}^{s} \mathbf{I}_N) \in \mathbb{C}^{N\times 1}$ and  $h_{d,k}^{m} \sim \mathcal{CN}(0,\beta_{d,k}^{m})\in \mathbb{C}^{1\times 1}$ are the  Rayleigh fading channels between user $k$ and the $s$'th RIS, and  user $k$ and the $m$'th AP, respectively, where $\beta_{2,k}^{s}$ and $\beta_{d,k}^{m}$ are the channel attenuation coefficients. Note that the subscripts $\{1,2,d\}$ denote the links, with $1$ denoting the link between an AP and an RIS, $2$ denoting the link between the RIS and a user, and $d$ denoting the direct link between the AP and a user. We use block fading model, where $h_{d,k}^{m}$ and $\mathbf{h}_{2,k}^{s}$ ($1 \leq m \leq M,1 \leq k \leq K, 1\leq s \leq S $) stay constant during a coherent interval and change independently between coherent intervals. Additionally, we assume channel reciprocity, i.e., the channel coefficients for uplink and downlink transmissions are the same. The response of the RIS is captured in $\boldsymbol{\Theta}=\text{diag}(\mathbf{v})\in \mathbb{C}^{N\times N}$, where $\mathbf{v}=[\alpha_1 \exp(j\theta_{1}), \alpha_2 \exp(j\theta_{2}),\dots, \alpha_N \exp(j\theta_{N})]^T\in \mathbb{C}^{N\times 1}$ is the RIS reflect beamforming vector, with $\theta_{n}\in[0,2\pi] $ being the induced phase-shift and $\alpha_{n} \in[0,1] $ being the given amplitude reflection coefficient of element $n$.  We denote components of $\mathbf{v}$ by $v_n=\alpha_n \exp(j\theta_n)$  in the remainder of this paper.\\
Given the above definitions, we can now define the total downlink channel at the user end, i.e., the channel between $m$'th AP to $k$'th user via all RISs, as $g_{m,k}=\sum_{s=1}^{S} \mathbf{h}_{1,m}^{s}\boldsymbol{\Theta} \mathbf{h}_{2,k}^{s}+h_{d,k}^{m}$, which is distributed as:
\begin{align}\label{eq1}
g_{m,k}\sim \mathcal{CN}(0,\rho_{m,k}),
\end{align}
where $\rho_{m,k}=\sum_{s=1}^{S}\beta_{2,k}^{s} \mathbf{h}_{1,m}^{s}\boldsymbol{\Theta}\boldsymbol{\Theta}^{H}\mathbf{h}_{1,m}^{s^{H}}+\beta_{d,k}^{m}$. In the following subsection we define channel estimates under the TDD protocol.
\section{Channel Estimation}\label{est_sec}
Under the TDD protocol, the AP $m$ exploits the channel reciprocity to estimate the downlink channels $g_{m,k}, k=1 \ldots K$  defined in \eqref{eq1}. As mentioned earlier, this paper assumes random phase-shift configuration at each RIS. Specifically, we assume $\alpha_{n}=1 $, $\forall n$, to achieve the largest array gain, and we choose $\theta_{n}$, $\forall n$ to be uniformly distributed between $[0, 2\pi]$. Under this setting, the APs only need to acquire the total channel $g_{m,k}$ estimate for each user $k$, without the need of estimating the individual channels $\mathbf{h}_{2,k}^{s}$,  $\forall s$, $\forall k$, and $h_{d,k}^{m}$, $\forall k$, which is challenging and time consuming \cite{nadeem, Q_w2}.\\ All users synchronously send pilot sequences $\boldsymbol{\phi}_{1} \ldots  \boldsymbol{\phi}_{K}$ to APs via the direct and RIS assisted links. The AP antennas then use the uplink training signal to acquire the channel estimates. Under the assumption that the number of users is less than the number of the orthogonal pilots, we assign mutually orthogonal pilot sequences to users, while assuming user mobility of less $10$ km/h and carrier frequency of $1.9$ GHz resulting in negligible pilot contamination \cite{Nayebi}. Let $\tau$ be the length of the coherence interval, $\tau_{c}$ be the length of uplink training duration per coherence interval, and $\tau_{d}$ be the length of downlink transmission duration (all in symbol-durations). It is required that $\tau_{c}<\tau$. Given the pilot sequence for each user $k$ is of length $\tau_c$ symbol-durations, $\boldsymbol{\phi}_{k} \in \mathbb{C}^{\tau_c \times 1}, k=1\ldots K$, where $||\boldsymbol{\phi}_{k}||^{2}=1$, then the $\tau_c \times 1$ signal received at the $m$'th AP is:
\begin{align}\label{eq2}
\mathbf{y}_{m}=\sqrt{\tau_{c} p_{c}} \sum_{k=1}^{K} g_{m,k} \boldsymbol{\phi}_{k}+ \mathbf{w}_{m},
\end{align}
where $p_{c}$ is the normalized transmit signal-to-noise ratio (SNR) of each pilot symbol and $\mathbf{w}_{m}$ is a vector of additive noise at the $m$'th AP whose elements are i.i.d. $\mathcal{CN}(0, 1)$ random variables (RVs).

Based on the received pilot signal in \eqref{eq2}, the $m$'th AP estimates the channel $g_{m,k}, k=1 \dots K$, by projecting $\mathbf{y}_{m}$ on $\boldsymbol{\phi}_{k}^{H}$, such that (s.t.) $\tilde{\mathbf{y}}_{m} \triangleq\boldsymbol{\phi}_{k}^{H}\mathbf{y}_{m}=\sqrt{\tau_{c}p_{c}}g_{m,k}+ \boldsymbol{\phi}_{k}^{H}\mathbf{w}_{m}$. Therefore, the minimum mean square estimation (MMSE) estimate of $g_{m,k}$ given $\tilde{\mathbf{y}}_{m}$ is defined as:
\begin{align}\label{eq4}
\hat{g}_{m,k}=\frac{\sqrt{\tau_{c}p_{c}}\rho_{m,k}\tilde{\mathbf{y}}_{m}}{\tau_{c}p_{c}\rho_{m,k}+1},
\end{align}
Let $\tilde{g}_{m,k}=g_{m,k}-\hat{g}_{m,k}$ be the channel estimation error. Since $\tilde{g}_{m,k}$ and $\hat{g}_{m,k}$ are uncorrelated \cite{Marz2}, we can statistically define the following:
\begin{align}
\hat{g}_{m,k}\sim \mathcal{CN}(0,\gamma_{m,k}),
\end{align}
where 
\begin{align}\label{eq_gamma}
\gamma_{m,k}=\frac{\tau_{c}p_{c}\rho_{m,k}\rho_{m,k}^{*}}{(\tau_{c}p_{c}\rho_{m,k}+1)},
\end{align}
and
\begin{align}
\tilde{g}_{m,k} \sim \mathcal{CN}(0, \rho_{m,k}-\gamma_{m,k}).
\end{align}
We aim at studying the achievable rate in the proposed RIS-CF-mMIMO system under this channel estimation protocol. In the following section, we investigate the conjugate precoding and derive the closed-form expression of the acheivable rate.

\section{Downlink Transmission}
In the downlink transmission, each AP treats the channel estimates as true channels and use the conjugate beamforming technique to transmit signals to $K$ users. Let $s_{k}$, $k= 1 \ldots K$, where $\mathbb{E}[|s_{k}|^{2}]=1$, be the symbol intended to user $k$, the transmit signal can then be expressed as follows:
\begin{align}
x_{m}=\sqrt{p_{d}}\sum_{k=1}^{K}\eta_{m,k}^{1/2}\hat{g}_{m,k}^{*}s_{k},
\end{align}
where $p_d$ is the total average power available at any AP, $\eta_{m,k}$, $m=1 \ldots M$, $k=1 \ldots K$, are power control coefficients chosen to satisfy the average power constraints at each AP, $\mathbb{E}[|x_{m}|^{2}]\leq p_{d}$. With the channel model defined in \eqref{eq1} and its estimate \eqref{eq4}, the power constraints can be rewritten as:
\begin{align}
\sum_{k=1}^{K}\eta_{m,k}\gamma_{m,k}\leq 1,
\end{align}
The received downlink signal at user $k$ can be defined as:
\begin{align}\label{rate1}
r_{k}&=\sum_{m=1}^{M}g_{m,k}x_{m}+n_{k}\\ \nonumber
&=\sqrt{p_{d}}\sum_{m=1}^{M}\sum_{k'=1}^{K}\eta_{m,k'}^{1/2}g_{m,k}\hat{g}_{m,k'}^{*}s_{k'}+ n_{k}
\end{align}
where $n_{k}$ represents additive Gaussian noise at the $k$'th user. We assume that $n_{k} \sim \mathcal{CN}(0, 1)$. Then $s_{k}$ will be decoded from $r_{k}$.
Now we define the closed-form expression of the achievable rate. Using similar analysis as in \cite{Hassibi, Marz1}, we assume that each user has the knowledge of signal statistics but not the channel realizations. Consequently, the received signal at the user $k$ can be written as:
\begin{align}\label{ach1}
r_{k}=\mathcal{D}_{k}s_{k}+\mathcal{B}_{k}s_{k}+\sum_{k'\neq k}^{K}\mathcal{U}_{kk'}s_{k'}+n_{k},
\end{align}
where $\mathcal{D}_{k}\delequal \sqrt{p_{d}}\mathbb{E}\left[{\sum}_{m=1}^{M}\eta_{m,k}^{1/2}g_{m,k}\hat{g}_{m,k}^{*}\right]$ is the deterministic value of the strength of the desired signal, $\mathcal{B}_{k}\delequal \sqrt{p_{d}}\left(\sum_{m=1}^{M}\eta_{m,k}^{1/2}g_{m,k}\hat{g}_{m,k}^{*}-\mathbb{E}\left[{\sum}_{m=1}^{M}\eta_{m,k}^{1/2}g_{m,k}\hat{g}_{m,k}^{*}\right]\right) $ is the beamforming gain uncertainty, and $U_{kk'}\delequal \sqrt{p_{d}}\sum_{m=1}^{M}\eta_{m,k'}^{1/2}g_{m,k}\hat{g}_{m,k'}^{*}$ is the interefence due to user $k'$.

It can be shown, through straightforward calculation, that the effective noise, $E_{k}=\left(\sqrt{p_{d}}\sum_{m=1}^{M}\sum_{k'=1}^{K}\eta_{m,k'}^{1/2}g_{m,k}\hat{g}_{m,k'}^{*}s_{k'}+ n_{k}\right)-\mathcal{D}_{k}s_{k}$, and the desired signal are uncorrelated. Knowing that uncorrelated Gaussian noise represents the worst case, we obtain the following achievable rate of the $k$'th user for the RIS-CF-mMIMO system: 
\begin{align}\label{ach2}
R_{k}=\log_{2}\left( 1+\frac{|\mathcal{D}_{k}|^{2}}{\mathbb{V}\text{ar}\{E_{k}\}}\right).
\end{align}
In the following theorem, we provide the closed-form expression of \eqref{ach2}, for a finite number of APs.

\begin{theorem}\label{Thm1} 
 The acheivable rate of transmission from the APs to the $k$'th user via all $S$ RISs is given by \eqref{ach_v} at the top of next page.
\begin{figure*}[ht]
\begin{align}
\label{ach_v}
{R}_{k}= \log_{2}\left(1+\frac{p_{d}\left(\sum_{m}^{M}\eta_{m,k}^{1/2}\gamma_{m,k}\right)^{2}}{p_{d}\sum_{k'=1}^{K}\sum_{m=1}^{M}\eta_{m,k'}\gamma_{m,k'}\rho_{m,k}+1}\right)
\end{align} 
\hrule
\end{figure*}
\end{theorem}

\begin{IEEEproof}
The proof is provided in Appendix A.
\end{IEEEproof}

From the expression in \eqref{ach_v}, we can notice the following.
The closed-form expression is a function of $\rho_{m,k}$ and $\gamma_{m,k}$, which are defined under a Rayleigh fading environment, while considering the estimation protocol proposed in Sec. \ref{est_sec}. Under the RIS response considered in Sec. \ref{est_sec}, where the amplitude $\alpha_{n}=1$, and $\theta_{n}$ is random, $\forall n$, we can define $\rho_{m,k}=\sum_{s=1}^{S}\beta_{2,k}^{s} \mathbf{h}_{1,m}^{s}\mathbf{h}_{1,m}^{s^{H}}+\beta_{d,k}^{m}$, since $\boldsymbol{\Theta}\boldsymbol{\Theta}^{H}=I_{N}$. Therefore, we can notice that under the considered channel fading and estimation method, both $\rho_{m,k}$ and $\gamma_{m,k}$ are independent of RIS phase-shifts $\theta_{n}$, $\forall n$. If an estimation protocol, such as ON-OFF protocol proposed in \cite{bayan}, were used, the RIS phase-shifts $\theta_{n}$ would appear in terms related to the direct channel estimation error. In such case, it is important to consider phase optimization. In our setting, however, we consider a sub-optimal scenario where the RISs are randomly configured. Therefore, it is reasonable to consider the derived acheivable rate as a lower bound to the RIS-assisted cell free performance metric.


\section{Numerical Results}\label{Sec:Sim}

In this section, we quantitively study the performance of RIS-CF-mMIMO and compare it to that of CF-mMIMO system. Specifically, we look into a CF-mMIMO system with $M$ APs and $K$ users uniformly distributed in a $D \times D$ $\text{km}^2$ geographical area. Under a similar setting, i.e., same number of uniformly distributed APs and user density, we additionally deploy $S$ uniformly distributed RISs, where each is equipped with $N$ reflecting elements to simulate the RIS-CF-mMIMO system. 

The large scale fading described in channel definition in \eqref{eq1} is a function of pathloss and shadow fading defined as follows \cite{Ngo}:
\vspace{-0.1 in}
\begin{align}
\beta^{t}_{l,w}=\mathcal{P}^{t}_{l,w} \cdot 10^{\frac{\sigma_{sh} z^{t}_{l,w}}{10}}
\end{align}
where $t \in [s,m]$, $l\in [1,2,d]$, and $w \in [m,k]$. The pathloss parameter $\mathcal{P}^{t}_{w,l}$ represents the pathloss between the $t$'th and $w$'th nodes via the $l$'th link, and $10^{\frac{\sigma_{sh}z^{t}_{l,w}}{10}}$ represents shadow fading with standard deviation $\sigma_{sh}$, and $z^{t}_{l,w}\sim \mathcal{N}(0,1)$. Let us denote the distance between the $t$'th node and the $w$'th node as $d_{t,w}$, where $t,w \in [m,s,k]$ for $m=1 \ldots M, s=1 \ldots S$ and $ k=1 \ldots K$. We use a three-slope model for the path loss \cite{Ngo}: In the case when $d_{t,w} >d_{1}$, we employ the Hata-COST231 propagation model, where the path loss exponent equals $3.5$ for ${h}_{d,k}^{m}$, $2.8$ for $\mathbf{h}_{2,k}^{s}$ and $2.0$ for $\mathbf{h}_{1,m}^{s}$\cite{Q_w2,bayan}. Otherwise, we set it equals to $2$ if $ d_{1} < d_{t,w}\leq d_{0}$, and equals $0$ if $d_{t,w} \leq d_{0}$. Note that when $d_{t,w} \leq d_{1}$ there is no shadowing \cite{Ngo}.

\begin{figure}[t]
\tikzset{every picture/.style={scale=.50}, every node/.style={scale=1.7}}
%
%
\definecolor{mycolor1}{rgb}{0.00000,0.44706,0.74118}%
\begin{tikzpicture}

\begin{axis}[%
width=11.108in,
height=4.789in,
at={(1.863in,0.734in)},
scale only axis,
xmin=50,
xmax=200,
xlabel={Number of APs ($M$)},
ymin=1,
ymax=3,
xtick={50, 75, 100, 125, 150, 175, 200},
ytick={1.2, 1.8, 2.4},
ylabel={Average sum rate (bps/Hz)},
x label style={font=\color{white!15!black}, at={(ticklabel cs:0.5)}},
y label style={font=\color{white!15!black}, at={(ticklabel cs:0.5)}},
axis background/.style={fill=white},
xmajorgrids,
ymajorgrids,
legend style={at={(0.03,1.45)}, anchor=south west, legend cell align=left, align=left, draw=white!15!black}
]
\addplot [color=mycolor1, line width=1.6pt, mark=asterisk,mark size=8.0pt, mark options={solid, mycolor1}]
  table[row sep=crcr]{%
50	1.19685341638829\\
100	1.84604983811438\\
150	2.28333686208531\\
200	2.62277782738982\\
};
\addlegendentry{User knows the effective channel gain}

\addplot [color=red, line width=1.5pt, mark=o,mark size=6.0pt, mark options={solid, red}]
  table[row sep=crcr]{%
50	1.17793759086239\\
100	1.81764867277164\\
150	2.2593005566073\\
200	2.59696475847386\\
};
\addlegendentry{User only knows the channel statistics }

\end{axis}
\end{tikzpicture}%
\caption{Validation of the closed form expression against the Monte-Carlo (MC) values, under imperfect CSI. Here we assume $D=1 \times 1 \text{km}^{2}$, $\beta_{l,w}^{t}=1$, $K=40, S=30$, and $N=10$.}
\label{equiv_model}
\end{figure}
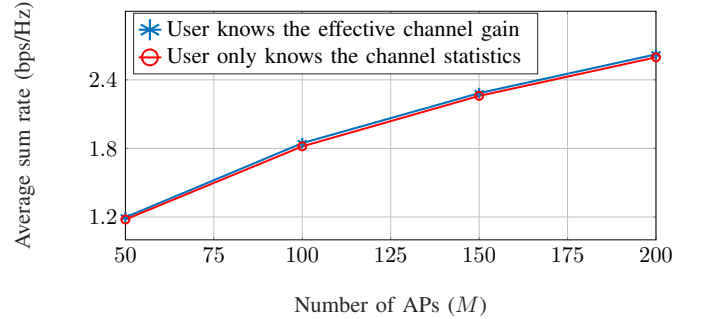

\begin{figure}[t]
\tikzset{every picture/.style={scale=.52}, every node/.style={scale=1.7}}
\input{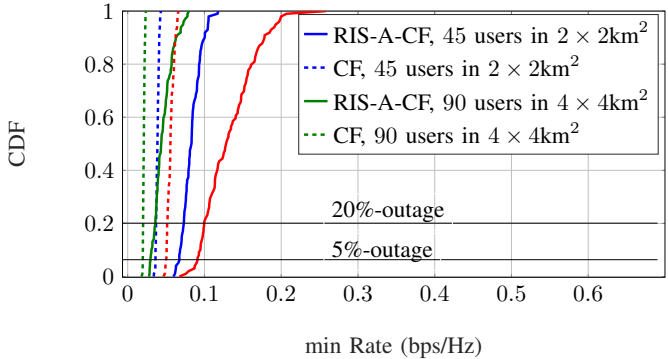}
\caption{The minimum rate for RIS-CF-mMIMO and CF-mMIMO under different $K$ and $D$ settings. Here, $M=100$ and $S=80$ and red curves are for $K=45$ and $D=4$.}
\label{fig3}
\end{figure}

In all our results, we use the parameters defined in Table \ref{T1}, unless otherwise noted. The quantities $\bar{p}_d$ and $\bar{p}_{c}$ in the table are the transmit powers for the transmit data and pilot symbol, respectively. The corresponding normalized transmit SNRs $p_{d}$ and $p_{c}$ can be computed by dividing these powers by the noise power, where the noise power is given by: noise power = bandwidth $\times k_B \times T_{0} \times$ noise figure (W), $k_B = 1.381 \times 10^ {-23}$ (Joule per Kelvin) is the Boltzmann constant, and $T_{0} = 290$ (Kelvin) is the noise temperature.
\begin{table}[t]\label{T1}
\centering
\caption{System Parameters}
\begin{tabular}{ |c|c|c| } 
\hline
 Parameter & Value \\
 \hline 
 carrier frequency & 1.9 GHz  \\ 
 \hline
 bandwidth & 20 MHz \\ 
 \hline
 noise figure & 9 dB \\ 
 \hline
 AP height & 15 m \\ 
 \hline
 RIS height & 18 m \\ 
  \hline
 user-device height & 1.65 m \\ 
  \hline
 $\sigma_{sh}$ & 8 dB \\ 
 \hline
 $\bar{p}_{d}$, $\bar{p}_c$  &  200 mW \cite{Nayebi} \\ 
  \hline
 $d_{0}$, $d_{1}$ & 10 m, 50 m \\ 
  \hline
 $\tau$ (coherence period) & 200 samples \\ 
 \hline
\end{tabular}
\label{T1}
\end{table}
In Fig. \ref{equiv_model}, we plot the closed-form expression of acheivable sum rate using \eqref{ach_v}, s.t.:
\begin{align}\label{sum_rate}
R_{sum}=(\frac{1-\tau_{c}/\tau}{2})\sum_{k=1}^{K}R_{k},
\end{align}
versus the varying number of APs $M$, along with the Monte-Carlo simulated curve. For the purpose of our simulation, we assume half the data samples are spent for downlink transmission. Specifically, we compare the sum rate under the achievable rate in \eqref{ach_v} with the rate in following expression:
\begin{align}\label{MC_rate}
\tilde{R}_{k}=\mathbb{E}\left[\log_{2}\left(1+\frac{p_{d}|\sum_{m=1}^{M} \eta_{m,k}^{1/2}g_{m,k}\hat{g}^{*}_{m,k}|^{2}}{p_{d}\sum_{k'\neq k}^{K}|\sum_{m=1}^{M}\eta_{m,k'}^{1/2}g_{m,k}\hat{g}^{*}_{m,k'}|^{2}+1}\right)\right]
\end{align}
which is the achievable rate under imperfect CSI at the users. Fig. \ref{equiv_model} shows the comparison between \eqref{ach_v}, which assumes that the users only know the channel statistics, and \eqref{MC_rate}, which assumes knowledge of the realizations. The small gap depicted in the plot proves the tightness of the expressions at a practical system dimension.
\newline We fix the number of reflecting elements per RIS in the following simulation figures to $N=30$. In Fig. \ref{fig3}, we compare the minimum rate, i.e., $\text{min}_{k}$ $R_{k}$, achieved for both RIS-CF-mMIMO and CF-mMIMO, to demonstrate the coverage enhancement under RIS deployment in the CF-mMIMO system. Specifically, the figure shows the cumulative distribution (CDF) of the minimum (min) rate for a CF-mMIMO with and without RIS deployment, under three different scenarios. We use the closed-form rate expression defined in \eqref{ach_v} for RIS-CF-mMIMO and the closed-form rate expression defined in Theorem 1 in \cite{Nayebi} for CF-mMIMO system. For all scenarios depicted in Fig. \ref{fig3}, we consider $M=100$ and $S=80$. Specifically, the blue curves examine the performance under $K=45$ users uniformly distributed in a $2 \times 2$ km$^2$ georaphical area, i.e., $D=2$, green curves are under $K=90$ users uniformly distributed in a $4 \times 4$ \text{km}$^2$, and the red curves are under $K=45$ users uniformly distributed in a $4 \times 4$ \text{km}$^2$ area.  

There are three important takeaways which we can draw from Fig. \ref{fig3}. First, looking at the blue curves (both dashed and solid) we can see that by deploying RISs, specifically $S=80$ and $N=30$ in this example, under a 5\% outage constraint, the minimum rate can be increased by a factor of $\approx$ $1.82$ as compared to CF-mMIMO without RIS. Second, an RIS-CF-mMIMO system serving double the number of users (green solid curve), $K=90$ in a $4 \times 4$ km$^2$ area (same user density but half AP density), can achieve an equivalent minimum rate as that of a CF-mMIMO serving $K=45$ in $2 \times 2$ km$^2$ area (blue dashed curve) under the same outage probability of 20\%. This implies that with the same number of AP deployment, we can double the network's service (serve more users) over a wider area. Again, note that our perfomance metric used, defined in \eqref{ach_v}, is a lower bound to the RIS-assisted performance. Therefore, the results in this work could be seen as a lower bound on the performance improvement that can be realized using an RIS in a CF-mMIMO system. Finally, the solid red curve looks at a scenario where an RIS-CF-mMIMO system serves $K=45$ users uniformly distributed in a $4 \times 4$ km$^2$ area and compares it with the CF-mMIMO system (red dashed line) serving users under a similar setting. Note that in this scenario, we have only spread the $45$ users over a wider area. This increases the average distance between the APs and users, which weakens AP-user link due to the induced pathloss values. The results show that the minimum rate is doubled by deploying RISs into the system under a 20\% outage constraint, and by a factor of about $1.87$  under a 5\% outage constraint. This shows that by deploying enough RISs, we can achieve better service for users suffering from weak AP-user links, such as in rural areas with low population.

\begin{figure}[t]
\tikzset{every picture/.style={scale=.478}, every node/.style={scale=1.75}}
\input{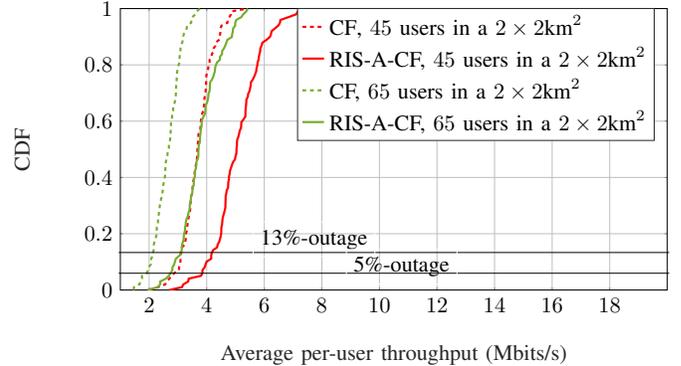}
\caption{The average per-user throughput for RIS-CF-mMIMO and CF-mMIMO under different $K$ settings. Here, $M=100$, $S=80$ and $D= 2$.}
\label{fig4}
\end{figure}

Next, we study the per-user throughput performance. Specifically, we consider the following per-user net throughput expression, which takes into account the channel estimation overhead \cite{Ngo}:
\begin{align}
S_{k}=B\frac{1-\frac{\tau_{c}}{\tau}}{2}R_{k},
\end{align}
where $B$ is the bandwidth. Fig. \ref{fig4} plots the average per-user throughput for RIS-CF-mMIMO (solid curves) and CF-mMIMO (dashed curves), under $K=45$ users (red curves) and $K=65$ users (green curves) uniformly distributed in a $2 \times 2$ $\text{km}^2$ area. From the red curves we can see that the average per-user throughput can be increased by almost a factor of $1.42$ in an RIS-CF-mMIMO system under a 5\% outage constraint. Additionally, as shown from the solid green curve, increasing the number of users to $65$ while deploying RISs, specifically $S=100$ and $N=30$ in this example, we can achieve the same throughput as that of CF-mMIMO with fewer users served per unit area, with an outage of 13\%.

\begin{figure}[t]
\tikzset{every picture/.style={scale=.47}, every node/.style={scale=1.8}}
%
%
\definecolor{mycolor1}{rgb}{0.07843,0.16863,0.54902}%
\definecolor{mycolor2}{rgb}{0.74902,0.00000,0.74902}%
\begin{tikzpicture}

\begin{axis}[%
width=12.529in,
height=6.452in,
at={(2.102in,0.871in)},
scale only axis,
xmin=80,
xmax=140,
xlabel style={font=\color{white!15!black},at={(ticklabel cs:0.5)}},
xlabel={M APs},
ymin=8.7,
ymax=12,
xtick={80, 90, 100, 110, 120, 130, 140},
ytick={9, 10, 11, 12},
ylabel style={font=\color{white!15!black},at={(ticklabel cs:0.5)}},
ylabel={Average sum rate (Bps/Hz)},
xmajorgrids,
ymajorgrids,
axis background/.style={fill=white},
legend style={at={(0.0099,1.42)}, anchor=south west, legend cell align=left, align=left, draw=white!15!black}
]
\addplot [color=red, dashed, line width=2pt, mark=square,mark size=5.0pt, mark options={solid, red}]
  table[row sep=crcr]{%
80	8.813017171\\
100	9.770895347\\
120	10.98367886\\
140	11.85389981\\
};
\addlegendentry{CF-mMIMO}

\addplot [color=mycolor1, line width=1.5pt, mark=star,mark size=7.0pt, mark options={solid, mycolor1}]
  table[row sep=crcr]{%
80	9.6696454243349\\
100	9.6696454243349\\
120	9.6696454243349\\
140	9.6696454243349\\
};
\addlegendentry{RIS-A-CF-mMIMO, M=70, N=30, S=80}

\addplot [color=mycolor2, line width=1.5pt, mark=star,mark size=7.0pt, mark options={solid, mycolor2}]
  table[row sep=crcr]{%
80	10.3415756519945\\
100	10.3415756519945\\
120	10.3415756519945\\
140	10.3415756519945\\
};
\addlegendentry{RIS-A-CF-mMIMO, M=70, N=30, S=200}

\end{axis}
\end{tikzpicture}%
\caption{The average sum rate comparison for RIS-CF-mMIMO and CF-mMIMO under different number of RISs $S$ deployment.}
\label{fig5}
\end{figure}
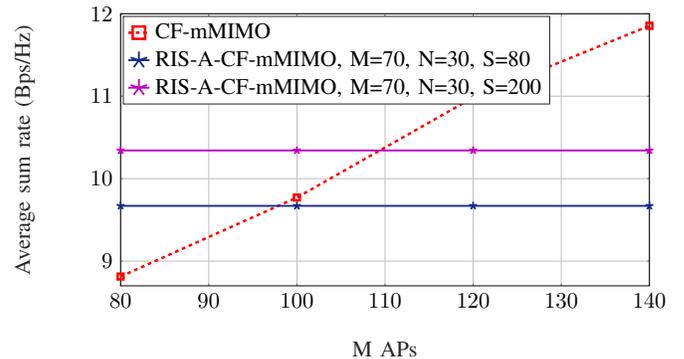

Finally, we look into a primitive approach of studying the effect of deploying RISs in CF-mMIMO on the sum-rate performance. Specifically, Fig. \ref{fig5} plots the average sum rate curve of CF-mMIMO over varying number of $M$ APs, while fixing the number of APs in the RIS-CF-mMIMO to $M=70$, and varying the number of RISs in $S \in \{80, 200\}$. Fig. \ref{fig5} shows that an RIS-CF-mMIMO system with $M=70$ APs and $S=80$ RISs can perform as well as a CF-mMIMO with $M=92$ APs, thus saving up to $22$ APs for an equivalent average sum rate performance. Additionally, an RIS-CF-mMIMO system with $M=70$ APs and $S=200$ RISs can perform as well as a CF-mMIMO with $M=107$ APs. Thus, again we can save up to $37$ APs by integrating low cost RISs into CF-mMIMO systems. Though we cannot draw a clear relationship between the number of RISs or reflecting elements per RIS that are needed per saved AP, we can still see a promising initial insight on resource saving when deploying RISs into a CF-mMIMO system. Further study is required to compute the potential number of RISs needed to deploy in CF-mMIMO to save an AP.

\section{Conclusion}
\label{Sec:Con}

This paper studied the performance of an RIS-CF-mMIMO under imperfect CSI. By deriving a closed-form expression of the acheivable rate, we show that under Rayleigh fading and a DE channel estimation protocol, the system's performance is independent of the RIS phase configuration. We further compare the performance of a RIS-CF-mMIMO system with the well known CF-mMIMO system. The results show that RISs deployment can improve the current CF-mMIMO topology in terms of system's coverage, nearly doubling the minimum rate relative to a CF-mMIMO system. Additionally, RIS technology can lead to a significant resource saving, where a number of low cost RISs may be deployed instead of a number of APs while achieving equivalent sum rate performance. A more accurate and qualitative comparison between the potential number of RIS reflecting elements/RIS surfaces needed to replace one AP under Rician fading channels and optimal RIS configuration will be studied as a future extension of this work.

\appendix

\subsection{Proof of Theorem 1} 
To derive the closed-form expression for the acheivable rate given in \eqref{ach_v}, we need to compute $\mathcal{D}_{k}$ and $\mathbb{V}\text{ar}\{E_{k}\}$. Recall that $E_{k}=\left(\sqrt{p_{d}}\sum_{m=1}^{M}\sum_{k'=1}^{K}\eta_{m,k'}^{1/2}g_{m,k}\hat{g}_{m,k'}^{*}s_{k'}+ n_{k}\right)-\mathcal{D}_{k}s_{k}$. It can be shown that $\mathcal{B}_{k}$, $\mathcal{U}_{kk'}$ and $n_{k}$ are uncorrelated, therefore we can define $\mathbb{V}\text{ar}\{E_{k}\}=\mathbb{E}[|\mathcal{B}_{k}|^{2}] + \sum_{k'\neq k}^{K}\mathbb{E}[|\mathcal{U}_{kk'}|^{2}] +1$.\\ 
Let $\mathcal{E}= g_{m,k}-\hat{g}_{m,k}$ be the channel estimation error. Then
\begin{align}
\mathcal{D}_{k}&=\sqrt{p_{d}}\mathbb{E}\left[{\sum}_{m=1}^{M}\eta_{m,k}^{1/2}g_{m,k}\hat{g}_{m,k}^{*}\right] \nonumber \\
&=\sqrt{p_{d}}\mathbb{E}\left[{\sum}_{m=1}^{M}\eta_{m,k}^{1/2}(\mathcal{E}+\hat{g}_{m,k})\hat{g}_{m,k}^{*}\right].\nonumber
\end{align}
Owing to the properties of MMSE estimation, $\mathcal{E}$ and $\hat{g}_{m,k}$ are independent, thus
\begin{align}\label{pr1}
\mathcal{D}_{k}&=\sqrt{p_{d}}{\sum}_{m=1}^{M}\eta_{m,k}^{1/2}\gamma_{m,k}.
\end{align}
Next, we compute $\mathbb{E}[|\mathcal{B}_{k}|^{2}]$. Recall that the variance of sum of independent RVs is equal to sum of the variances, therefore, 
\begin{align}
\mathbb{E}[|\mathcal{B}_{k}|^{2}]&=p_{d}\sum_{m=1}^{M}\eta_{m,k}\mathbb{E}\left[|g_{m,k}\hat{g}_{m,k}^{*}-\mathbb{E}\left[g_{m,k}\hat{g}_{m,k}^{*}\right]|^{2}\right] \nonumber \\
=&p_{d}\sum_{m=1}^{M}\eta_{m,k}\left(\mathbb{E}\left[|g_{m,k}\hat{g}_{m,k}^{*}|^{2}\right]-|\mathbb{E}\left[g_{m,k}\hat{g}_{m,k}^{*}\right]|^{2}\right)\nonumber \\
=&p_{d}\sum_{m=1}^{M}\eta_{m,k}\left(\mathbb{E}\left[\left|\mathcal{E}\hat{g}_{m,k}^{*}+|\hat{g}_{m,k}|^{2}\right|^{2}\right]-\gamma_{m,k}^{2}\right). \nonumber
\end{align}
Following similar argument in the proof of Theorem 1 provided in \cite{Ngo}, it can be shown that the above can be simplified to:
\vspace{-0.1 in}
\begin{align}\label{pr2}
\mathbb{E}[|\mathcal{B}_{k}|^{2}]=p_{d}\sum_{m=1}^{M}\eta_{m,k}\gamma_{m,k}\rho_{m,k}.
\end{align}
Finally, to compute $\mathbb{E}[|\mathcal{U}_{kk'}|^{2}]$, note that
\begin{align}\label{pr44}
\mathbb{E}[|\mathcal{U}_{kk'}|^{2}]=p_{d}\mathbb{E}\left[|\sum_{m=1}^{M}\eta_{m,k'}^{1/2}g_{m,k}\hat{g}_{m,k'}^{*}|^{2}\right].
\end{align}
Recall that ($g_{m',k}$, $g_{m,k}$), and ($g_{m,k'}$,$g_{m,k}$), for $m' \neq m$ and $k' \neq k$, are independent. By expanding the expression inside the expectation and following a similar proof as that of Theorem 1 in \cite{Ngo}, while considering our setting, i.e., orthogonal pilots, (omitted here due to lack of space) we can express \eqref{pr44} as:
\begin{align} \label{pr3}
\mathbb{E}[|\mathcal{U}_{kk'}|^{2}]=p_{d}\sum_{m=1}^{M}\eta_{m,k'}\rho_{m,k}\gamma_{m,k'}.
\end{align}
Plugging \eqref{pr3} and \eqref{pr2} in the definition $\mathbb{V}\text{ar}\{E_{k}\}$, and using the definition and \eqref{pr1} back in \eqref{ach_v} completes the proof.
\bibliographystyle{IEEEtran}
\bibliography{bibl}

\begin{thebibliography}{10}
\providecommand{\url}[1]{#1}
\csname url@samestyle\endcsname
\providecommand{\newblock}{\relax}
\providecommand{\bibinfo}[2]{#2}
\providecommand{\BIBentrySTDinterwordspacing}{\spaceskip=0pt\relax}
\providecommand{\BIBentryALTinterwordstretchfactor}{4}
\providecommand{\BIBentryALTinterwordspacing}{\spaceskip=\fontdimen2\font plus
\BIBentryALTinterwordstretchfactor\fontdimen3\font minus
  \fontdimen4\font\relax}
\providecommand{\BIBforeignlanguage}[2]{{%
\expandafter\ifx\csname l@#1\endcsname\relax
\typeout{** WARNING: IEEEtran.bst: No hyphenation pattern has been}%
\typeout{** loaded for the language `#1'. Using the pattern for}%
\typeout{** the default language instead.}%
\else
\language=\csname l@#1\endcsname
\fi
#2}}
\providecommand{\BIBdecl}{\relax}
\BIBdecl

\bibitem{Ngo}
H.~Q. {Ngo}, A.~{Ashikhmin}, H.~{Yang}, E.~G. {Larsson}, and T.~L. {Marzetta},
  ``Cell-free massive mimo versus small cells,'' \emph{IEEE Transactions on
  Wireless Communications}, vol.~16, no.~3, pp. 1834--1850, 2017.

\bibitem{Zhang}
J.~{Zhang}, S.~{Chen}, Y.~{Lin}, J.~{Zheng}, B.~{Ai}, and L.~{Hanzo},
  ``Cell-free massive mimo: A new next-generation paradigm,'' \emph{IEEE
  Access}, vol.~7, pp. 99\,878--99\,888, 2019.

\bibitem{Nayebi}
E.~{Nayebi}, A.~{Ashikhmin}, T.~L. {Marzetta}, and H.~{Yang}, ``Cell-free
  massive mimo systems,'' in \emph{2015 49th Asilomar Conference on Signals,
  Systems and Computers}, 2015, pp. 695--699.

\bibitem{QNgo}
H.~Q. Ngo, A.~Ashikhmin, H.~Yang, E.~G. Larsson, and T.~L. Marzetta,
  ``Cell-free massive mimo: Uniformly great service for everyone,'' 2015.

\bibitem{Chen}
Z.~{Chen} and E.~{Bjoernson}, ``Can we rely on channel hardening in cell-free
  massive mimo?'' in \emph{2017 IEEE Globecom Workshops (GC Wkshps)}, 2017, pp.
  1--6.

\bibitem{Chen2}
Z.~Chen and E.~Björnson, ``Channel hardening and favorable propagation in
  cell-free massive mimo with stochastic geometry,'' 2018.

\bibitem{AAIB}
A.~A.~I. {Ibrahim}, A.~{Ashikhmin}, T.~L. {Marzetta}, and D.~J. {Love},
  ``Cell-free massive mimo systems utilizing multi-antenna access points,'' in
  \emph{2017 51st Asilomar Conference on Signals, Systems, and Computers},
  2017, pp. 1517--1521.

\bibitem{haung}
C.~Huang, Member, IEEE, R.~Mo, C.~Yuen, and S.~Member, ``Reconfigurable
  intelligent surface assisted multiuser miso systems exploiting deep
  reinforcement learning,'' 2020.

\bibitem{Marco}
M.~D. Renzo, M.~Debbah, D.-T. Phan-Huy, A.~Zappone, M.-S. Alouini, C.~Yuen,
  V.~Sciancalepore, G.~C. Alexandropoulos, J.~Hoydis, H.~Gacanin, J.~de~Rosny,
  A.~Bounceu, G.~Lerosey, and M.~Fink, ``Smart radio environments empowered by
  ai reconfigurable meta-surfaces: An idea whose time has come,'' 2019.

\bibitem{Gong}
S.~{Gong}, X.~{Lu}, D.~T. {Hoang}, D.~{Niyato}, L.~{Shu}, D.~I. {Kim}, and
  Y.~C. {Liang}, ``Towards smart wireless communications via intelligent
  reflecting surfaces: A contemporary survey,'' \emph{IEEE Communications
  Surveys Tutorials}, pp. 1--1, 2020.

\bibitem{pan}
C.~Pan, H.~Ren, K.~Wang, M.~Elkashlan, M.~Chen, M.~D. Renzo, Y.~Hao, J.~Wang,
  A.~L. Swindlehurst, X.~You, and L.~Hanzo, ``Reconfigurable intelligent
  surface for 6g and beyond: Motivations, principles, applications, and
  research directions,'' 2020.

\bibitem{bayan}
B.~Al-Nahhas, Q.-U.-A. Nadeem, and A.~Chaaban, ``Intelligent reflecting surface
  assisted miso downlink: Channel estimation and asymptotic analysis,'' 2020.

\bibitem{nadeem}
Q.-U.-A. Nadeem, A.~Kammoun, A.~Chaaban, M.~Debbah, and M.-S. Alouini,
  ``Intelligent reflecting surface assisted wireless communication: Modeling
  and channel estimation,'' 2019.

\bibitem{secure}
Z.~{Chu}, W.~{Hao}, P.~{Xiao}, and J.~{Shi}, ``Intelligent reflecting surface
  aided multi-antenna secure transmission,'' \emph{IEEE Wireless Communications
  Letters}, vol.~9, no.~1, pp. 108--112, 2020.

\bibitem{noma}
X.~Yue and Y.~Liu, ``Performance analysis of intelligent reflecting surface
  assisted noma networks,'' 2020.

\bibitem{UAV}
D.~Ma, M.~Ding, and M.~Hassan, ``Enhancing cellular communications for uavs via
  intelligent reflective surface,'' 2020.

\bibitem{2020decentralized}
S.~Huang, Y.~Ye, M.~Xiao, H.~V. Poor, and M.~Skoglund, ``Decentralized
  beamforming design for intelligent reflecting surface-enhanced cell-free
  networks,'' 2020.

\bibitem{2020capacity}
Z.~Zhang and L.~Dai, ``Capacity improvement in wideband reconfigurable
  intelligent surface-aided cell-free network,'' 2020.

\bibitem{2020energy}
Q.~N. Le, V.-D. Nguyen, and O.~A. Dobre, ``Energy efficiency maximization in
  ris-aided cell-free network with limited backhaul,'' 2020.

\bibitem{Q_w2}
Q.~{Wu} and R.~{Zhang}, ``Intelligent reflecting surface enhanced wireless
  network via joint active and passive beamforming,'' \emph{IEEE Transactions
  on Wireless Communications}, vol.~18, no.~11, pp. 5394--5409, 2019.

\bibitem{Marz2}
J.~{Hoydis}, S.~{ten Brink}, and M.~{Debbah}, ``Massive mimo in the ul/dl of
  cellular networks: How many antennas do we need?'' \emph{IEEE Journal on
  Selected Areas in Communications}, vol.~31, no.~2, pp. 160--171, 2013.

\bibitem{Hassibi}
B.~{Hassibi} and B.~M. {Hochwald}, ``How much training is needed in
  multiple-antenna wireless links?'' \emph{IEEE Transactions on Information
  Theory}, vol.~49, no.~4, pp. 951--963, 2003.

\bibitem{Marz1}
H.~{Yang} and T.~L. {Marzetta}, ``Capacity performance of multicell large-scale
  antenna systems,'' in \emph{2013 51st Annual Allerton Conference on
  Communication, Control, and Computing (Allerton)}, 2013, pp. 668--675.

\end{thebibliography}

\end{document}